\documentclass[journal]{IEEEtran}

\makeatother
\usepackage{amssymb,amsmath}
\usepackage{cite}
\usepackage{color,graphicx}

\newtheorem{secthm}{Theorem}[section]
\newtheorem{seccor}[secthm]{Corollary}

\newtheorem{secex}[secthm]{Example}
\newtheorem{secprop}[secthm]{Proposition}
\newtheorem{secdefn}[secthm]{Definition}
\newtheorem{secrem}[secthm]{Remark}
\newtheorem{secasm}[secthm]{Assumption}

\newcommand{\cE} { {\cal E}}
\newcommand{\bR} { {\mathbb R}}

\def\red{\hfill $\lhd$}

\begin{document}
\title{Krasovskii and Shifted Passivity Based Control}
\author{
Yu Kawano, Krishna Chaitanya Kosaraju, and Jacquelien M. A. Scherpen
\thanks{Y. Kawano is with the Graduate School of Advanced Science and Engineering, Hiroshima University, Japan (email: ykawano@hiroshima-u.ac.jp).}
\thanks{K.C. Kosaraju is with the Department of Electrical Engineering, University of Notre Dame, Notre Dame, IN 46556, USA (email: kkosaraj@nd.edu).}
\thanks{J.M.A. Scherpen is with the Jan C. Willems Center for Systems and Control and the Faculty of Science and Engineering, University of Groningen, Groningen, The Netherlands (email: j.m.a.scherpen@rug.nl).}
\thanks{This work of Kawano was supported in part by JSPS KAKENHI Grant Number JP19K23517.}
\thanks{This work of Kosaraju and Scherpen was supported in part by the Netherlands Organisation for Scientific Research through Research Programme ENBARK+ under Project 408.urs+.16.005.}
\thanks{\color{black}The preliminary version of the paper was presented at the 11th IFAC symposium on nonlinear control systems in 2019~\cite{KKS:19}.}}
 
\maketitle

\IEEEpeerreviewmaketitle

\begin{abstract}
In this paper, our objective is to develop novel passivity based control techniques by introducing a new passivity concept named {\it Krasovskii passivity}. As a preliminary step, we investigate properties of Krasovskii passive systems and establish relations among four relevant passivity concepts including Krasovskii passivity. Then,  we develop novel dynamic controllers based on Krasovskii passivity and based on shifted passivity. 
\end{abstract}


\section{Introduction}
 Passivity as a \emph{tool} enables us to develop various types of passivity based control (PBC) techniques, and moreover as a \emph{property}, it helps us to understand these techniques in the standard engineering parlance. Lyapunov analysis discusses stability with respect to an equilibrium. However, notions like differential (or called contraction) analysis and incremental stability~\cite{Angeli:02,LS:98,FS:14,KBC:20,TRK:18,AJP:16,KS:16} study the convergence between any pair of trajectories. These different notions have resulted in diverse stability definitions, which further resulted in disparate passivity definitions such as differential passivity and incremental passivity~\cite{PM:08,Schaft:13,FSS:13,Schaft:00}. 

There are several papers that describe these relatively new differential passivity concepts~\cite{Schaft:13,FSS:13}. Apart from the elegance of analysis, it is not well understood how differential passivity can be used either as a tool or as a property although there is a few differential passivity based control techniques~\cite{KCS:18,KCP:18,CLK:19,RDS:18}. This is because, generally, differential passivity can be interpreted as a property of the variational system and does not give direct conclusions for the original system itself. 

In contrast, incremental passivity has been used as a tool or as a property via so called shifted passivity~\cite{JOG:07,OSM:01,Simpson:18}. If a system has an equilibrium point, incremental passivity results into shifted passivity at the equilibrium point. Shifted passivity can be interpreted as a generalization of standard passivity for a system whose equilibrium point is not necessarily the origin and is applied to various situations, see, e.g.~\cite{JOG:07,OSM:01,Simpson:18}. 
 
Similarly, for differential passivity it can be of interest to consider a passivity property at an equilibrium point that can help with analysis and control design. Motivated by this, we establish a new passivity concept, which we call \emph{Krasovskii passivity}. First, we marshal the aforementioned relevant four passivity concepts: 1) differential passivity, 2) Krasovskii passivity, 3) incremental passivity, and 4) shifted passivity. Then we establish a similar connection as the connection between incremental passivity and shifted passivity, for differential passivity and Krasovskii passivity. Furthermore, we show that Krasovskii passivity implies shifted passivity, and differential passivity with respect to a constant metric implies incremental passivity. Next, we provide novel dynamic control techniques based on Krasovskii passivity, which also inspire a new shifted passivity based dynamic controller. The utility of the proposed controllers are illustrated by a DC-Zeta converter. It is worth mentioning that to the best of our knowledge for this converter, a passivity based controller has not been designed in literature. 

In the preliminary conference version~\cite{KKS:19}, we have proposed Krasovskii passivity, provided sufficient conditions for port-Hamiltonian and gradient systems to be Krasovskii passive, and gave a brief introduction of Krasovskii passivity based control techniques. However, the considered storage function is restricted into one with a constant metric. Also, incremental passivity and shifted passivity have not been considered in the preliminary version. This paper contains the following new contributions:
\begin{itemize}
\item general storage functions are used for analysis and controller design;
\item a necessary and sufficient condition for Krasovskii passivity is presented;
\item we establish relations among four types of passivity properties;
\item we show an example of a Krasovskii passive system, which is not differentially passive with a positive definite storage function;
\item the proposed Krasovskii passivity based dynamic controller is more general than the one in \cite{KKS:19}.
\item we newly present a shifted passivity based dynamic controller.
\end{itemize}

The remainder of this paper is organized as follows. In Section~\ref{sec:PP}, we define Krasovskii passivity and establish the connection among the four passivity concepts. In Section~\ref{sec:PBC}, we design two novel passivity based dynamic controllers on the basis of Krasovskii passivity and of shifted passivity. In Section~\ref{sec:Ex}, the two provided controllers are applied to solve the stabilization problem of a DC-Zeta converter. Finally, Section~\ref{sec:con} concludes this paper.

{\it Notation:}
The set of real numbers and non-negative real numbers are denoted by $\bR$ and $\bR_+$, respectively. For symmetric matrices $P,Q \in \bR^{n\times n}$, $P \succ Q$ ($P \succeq Q$) means that $P-Q$ is positive definite (semidefinite). For a vector $x \in \bR^n$, define~$|x|_P:= \sqrt{x^\top P x}$, where~$P\in \bR^{n \times n}$. If~$P$ is the identity matrix, this is nothing but the Euclidean norm and is simply denoted by~$|x|$. The open ball of radius~$r > 0$ centered at~$p \in \bR^n$ is denoted by~$B_r(p):=\{x \in \bR^n: |x-p| < r\}$; its dimension ($n$ in this case) is not stated explicitly because this is clear from the context.  

\section{Analysis of Passivity Properties}\label{sec:PP}
\subsection{Preliminaries}
Consider the following input-affine nonlinear system:
\begin{align}\label{eq:sys}
\left\{\begin{array}{l}
\displaystyle \dot x=f(x,u):=g_0(x)+\sum_{i=1}^{m}g_i(x)u_i,\\
y = h(x)
\end{array}\right.
\end{align}
where~$x:\bR_+ \to \bR^n$,~$u:\bR_+ \to  \bR^m$, and~$y: \bR_+ \to \bR^m$ denote the state, input, and output, respectively. Functions~$g:\bR^n \to \bR^{n \times m}$ and~$h:\bR^n \to \bR^m$ are of class~$C^1$, where~$u_i, y_i$ denote the~$i$th elements of~$u$ and~$y$, respectively, and $g_i$ denotes the $i$th column of $g$.

In some of our developments, we assume an equilibrium point to exist, i.e., we use the following assumption. 
\begin{secasm}\label{asm:steady-state}
For the system~\eqref{eq:sys}, the following set 
\begin{align*}
\cE :=\{(x^*,u^*) \in \bR^n \times \bR^m:f(x^*,u^*)=0\}
\end{align*}
is not empty. \red
\end{secasm}

In this paper, our objective is to develop new passivity based control (PBC) techniques by investigating different passivity properties from the standard one. To be self-contained, we first summarize results on standard passivity, which will be extended to passivity concepts defined in this paper.
\begin{secdefn}\label{def:dissipative}\cite{Schaft:00,Khalil:96}
The system~\eqref{eq:sys} is said to be \emph{passive} if there exists a class~$C^1$ function~$S:\bR^n\to \bR_+$, called the \emph{storage function}, such that~$S(0)=0$ and
\begin{align*}
\frac{\partial S(x)}{\partial x} f(x,u)  \le u^\top y
\end{align*}
for all~$(x,u) \in \bR^n \times \bR^m$.
\red
\end{secdefn}

The following necessary and sufficient condition is well known for standard passivity.
\begin{secprop}\label{prop:passivity}\cite[Corollary 4.1.5]{Schaft:00}
A system~\eqref{eq:sys} is passive if and only if there exists a class~$C^1$ function~$S:\bR^n\to \bR_+$ such that~$S(0) = 0$ and
\begin{align*}
&\frac{\partial S(x)}{\partial x} g_0(x)  \le 0,\\
&\frac{\partial S(x)}{\partial x} g (x) = h^\top(x)
\end{align*}
for all~$x \in \bR^n$.
\red
\end{secprop}
\begin{secrem}
For passivity analysis, $S(0)=0$ is not required in general; see e.g.~\cite{Schaft:00}. However, $S(0)=0$, $f(0,0)=0$, and $h(0)=0$ are standard assumptions for PBC design. We will impose similar assumptions for the ease of discussions. \red
\end{secrem}

For a passive system satisfying $f(0,0)=0$ and $h(0)=0$, the controller~$u = -K y$,~$K \succ 0$ plays an important role. According to~\cite{Schaft:00,Khalil:96}, this controller achieves stabilization of the origin if~$S(x)$ is positive definite, and a passive system~\eqref{eq:sys} is zero-state detectable in the following sense.
\begin{secdefn}
Suppose that Assumption~\ref{asm:steady-state} holds, and~$h(x^*)=0$. The system~\eqref{eq:sys} is said to be \emph{detectable} at~$(x^*,u^*) \in \cE$ if 
\begin{align*}
u(\cdot) = u^* \mbox{ and } y(\cdot )=0 \implies \lim_{t\to\infty} x(t) =x^*.
\end{align*}
If $f(0,0)=0$ and $h(0)=0$, detectability at~$(0,0)$ is called zero-state detectability~\cite{Schaft:00,Khalil:96}.
\red
\end{secdefn}

\begin{figure}[t]
\begin{center}
\includegraphics[width=80mm]{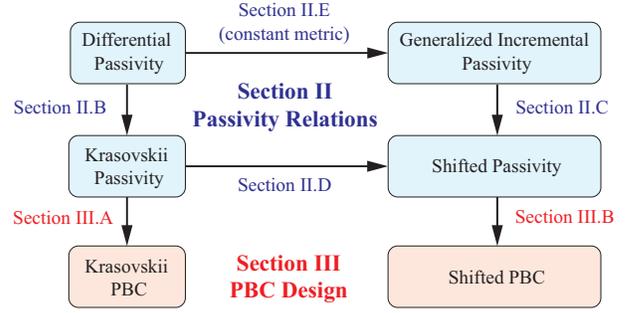}
\caption{Relationships among proposed passivity concepts and controllers.}
\label{fig:passivity}
\end{center}
\end{figure}

The rest of this section is dedicated to provide four passivity concepts and investigate their relations. These relations are summarized in Fig.~\ref{fig:passivity}. 

\subsection{Differential Passivity and Krasovskii Passivity}\label{sec:DPKP}
The differential passivity~\cite{Schaft:13} is introduced by using the so-called prolonged system consisting of the nonlinear system~\eqref{eq:sys} and its variational system:
\begin{align}
 \left\{\begin{array}{l}
\displaystyle\dot{\delta x} = F(x,u) \delta x+ \sum_{i=1}^m g_i(x)\delta u_i,\label{eq:vsys}\\
\displaystyle \delta y = h_d(x) \delta x,
\end{array}\right.
\end{align}
where~$\delta x:\bR_+ \to \bR^n$,~$\delta u:\bR_+ \to  \bR^m$, and~$\delta y :\bR_+ \to \bR^m$ denote the state, input, and output of the variational system, respectively, and
\begin{align*}
F(x,u):=\dfrac{\partial g_0(x)}{\partial x}+\sum_{i=1}^{m}\dfrac{\partial g_i(x)}{\partial x}u_i.
\end{align*}
The function~$h_d:\bR^n \to \bR^{m \times n}$ is continuous. Note that we do not assume that~$h_d(x)$ is~$\partial h(x)/\partial x$ in this paper.

Differential passivity is defined as a passivity property of the prolonged system.
\begin{secdefn}\cite{Schaft:13}
The nonlinear system~\eqref{eq:sys} is said to be \emph{differentially passive} if there exists a class~$C^1$ function~$S_D:\bR^n \times \bR^n\to \bR_+$ such that~$S_D(x,0)=0$ and
\begin{align}
&\frac{\partial S_D(x,\delta x)}{\partial x} f(x,u) + \frac{\partial S_D(x,\delta x)}{\partial \delta x} ( F(x,u)\delta x + g(x) \delta u)\nonumber\\
& \le \delta u^\top \delta y
\label{DPdef:eq}
\end{align}
for all~$(x,u) \in \bR^n \times \bR^m$ and~$(\delta x, \delta u)  \in \bR^n \times \bR^m$.
\red
\end{secdefn}

By applying Proposition~\ref{prop:passivity} to the prolonged system, its necessary and sufficient condition is obtained as follows.
\begin{secprop}\label{prop:DP}
A system~\eqref{eq:sys} is differentially passive if and only if there exists a class~$C^1$ matrix-valued function~$M(x) \succeq 0$,~$x \in \bR^n$ such that~$S_D:= (\delta x^\top M(x) \delta x)/2$ satisfies
\begin{align}
&\frac{\partial S_D(x,\delta x)}{\partial x} f(x,u) + \frac{\partial S_D(x,\delta x)}{\partial \delta x} F(x,u)\delta x \le 0,\label{cond:DP}\\\
&\frac{\partial S_D(x,\delta x)}{\partial \delta x} g(x) = (h_d(x) \delta x)^\top \nonumber
\end{align}
for all~$(x,u) \in \bR^n \times \bR^m$ and~$\delta x \in \bR^n$.
\end{secprop}
\begin{IEEEproof}
According to \cite[Proposition~4.2]{Schaft:13}, the system is differentially passive if and only if there exists a class~$C^1$ function~$S_D : \bR^n \times \bR^n \to \bR_+$ such that the above two conditions hold. Next, the variational system is linear with respect to~$(\delta x, \delta u, \delta y)$, and it is known that for passive linear systems, the storage function is a quadratic function~\cite[Theorem 14.2]{Terrell:09}. Therefore, if the system is differentially passive, the storage function can be described as~$S_D= (\delta x^\top M(x) \delta x)/2$ with~$M(x) \succeq 0$.
\end{IEEEproof}

It is worth mentioning that for differentially passive systems, a control design methodology has not been well explored yet. A bottle neck is that if one simply applies PBC techniques, then a controller is designed for the variational system, but not for the original system. It is not obvious how to design a controller for the original system from one designed for the variational system. To address this issue, we provide a new passivity concept, which we call \emph{Krasovskii passivity}; the reason for picking this name will be explained in Remark~\ref{rem:KP}. In fact, we will show that every differentially passive system is Krasovskii passive. This fact is helpful when one considers stabilizing controller design for differentially passive systems.

The main idea of defining Krasovskii passivity comes from the fact that the pair~$(f(x,u), \dot u)$ satisfies the equation of the variational system~\eqref{eq:vsys}, namely
\begin{align}\label{eq:idea_KP}
\frac{d f(x,u)}{dt} =F(x,u) f(x,u) + g(x) \dot u
\end{align}
Therefore, it is expected that if a system is differentially passive with the input and output port variables,~$\delta u$ and~$\delta y$, then it is passive for~$\dot u$ and~$h_d(x) f(x,u)$. To obtain this conclusion formally, we introduce the so called extended system~\cite{Schaft:82}:
\begin{align}\label{eq:esys}
\left\{\begin{array}{l}
\dot x = f(x,u),\\
\dot u = u_K,\\[1mm]
\displaystyle y_K =  h_K(x,u),
\end{array}\right.
\end{align}
where~$(x,u):\bR_+ \to \bR^n \times \bR^m$,~$u_K:\bR_+ \to \bR^m$, and~$y_K:\bR_+ \to \bR^m$ are the state, input, and output of the extended system, respectively, and~$h_K:\bR^n \times \bR^m \to \bR^m$ is continuous.

We now define Krasovskii passivity as follows.
\begin{secdefn}
Suppose that Assumption~\ref{asm:steady-state} holds. Then, the system~\eqref{eq:sys} is said to be \emph{Krasovskii passive} at~$(x^*,u^*) \in \cE$ if there exists a class~$C^1$ function~$S_K:\bR^n \times \bR^m \to \bR_+$ such that~$S_K(x^*,u^*)=0$ and
\begin{align}
\frac{\partial S_K(x,u)}{\partial x} f(x,u) + \frac{\partial S_K(x,u)}{\partial u} u_K \le u_K^\top y_K
\label{passive:eq}
\end{align}
for all~$(x,u) \in \bR^n \times \bR^m$ and~$u_K \in \bR^m$.
\red
\end{secdefn}

As expected, differential passivity implies Krasovskii passivity. However, the converse is not always true for positive definite storage functions as shown in Example~\ref{ex:KP} below.
\begin{secthm}\label{thm:DPKP}
Under Assumption~\ref{asm:steady-state}, if a system~\eqref{eq:sys} is differentially passive, then it is Krasovskii passive at any~$(x^*,u^*) \in \cE$ for~$h_K(x,u)=h_d(x)f(x,u)$.
\end{secthm}
\begin{IEEEproof}
By direct computation, it is possible to show that~$S_K(x,u)=S_D(x,f(x,u))$ is a storage function for Krasovskii passivity if $S_D(x,\delta x)$ is a storage function for differential passivity.
\end{IEEEproof}
\begin{secrem}\label{rem:KP}
In the proof of Theorem~\ref{thm:DPKP}, we obtain the storage function~$S_K(x,u)$ by replacing $\delta x$ by $f(x,u)$ in $S_D(x,\delta x)$. In fact, for stability analysis, the Lyapunov function constructed by Krasovskii's method is obtained similarly from the differential Lyapunov function of contraction analysis~\cite{FS:14}. Because of this analogy, we name the proposed passivity concept Krasovskii passivity.
\red
\end{secrem} 
\begin{secex}\label{ex:KP}
In this example, we show that Krasovskii passivity does not imply differential passivity with respect to a positive definite storage function.
Consider the following system: 
\begin{align*}
\left\{\begin{array}{l}
\begin{bmatrix}
\dot x_1 \\ \dot x_2
\end{bmatrix}
= g(x) u, \;
g(x) :=
\begin{bmatrix}
x_1^{1/3} \\ -x_2^{1/3}
\end{bmatrix},\\
\hspace{5.5mm}y = h(x) := x_1^{4/3} - x_2^{4/3},\\[2mm]
\displaystyle \hspace{3mm}y_K = \dot y =  \frac{\partial h(x)}{\partial x} g(x) u
\end{array}\right.
\end{align*}
where
\begin{align*}
\cE = &\{ (x^*,u^*) \in \bR^2 \times \bR : x^* =0 \}\\
& \cup \{ (x^*,u^*) \in \bR^2 \times \bR : u^* =0 \}.
\end{align*}
First, we show that this system is Krasovskii passive with respect to the following storage function:
\begin{align*}
S_K(x,u)=(1/2)|g(x)u|^2 = (x_1^{2/3}+x_2^{2/3}) u^2/2,
\end{align*}
where~$S_K(\cdot,\cdot) \ge 0$ and~$S_K(x^*,u^*)=0$ for any~$(x^*,u^*) \in \cE$. Compute
\begin{align*}
&\frac{\partial S_K(x,u)}{\partial x} g(x) u = 0,\\
&\frac{\partial S_K(x,u)}{\partial u}  = (x_1^{2/3}+x_2^{2/3}) u = \frac{\partial h(x)}{\partial x} g(x) u.
\end{align*}
Therefore, Proposition~\ref{prop:KP} implies that the system is Krasovskii passive at any~$(x^*,u^*) \in \cE$.

Next, we show that the system does not admit a storage function for differential passivity in the form of~$S_D(x,\delta x)=(\delta x^\top M(x) \delta x)/2$ with positive definite~$M(\cdot) \succ 0$. This can be done by showing non existence of such a storage function satisfying~\eqref{cond:DP}, i.e.,
\begin{align*}
&\frac{u}{2} \delta x^\top \left( \frac{\partial M(x)}{\partial x_1} x_1^{1/3} - \frac{\partial M(x)}{\partial x_2} x_2^{1/3} \right) \delta x\\
&+ \frac{u}{3}  \delta x^\top M(x) 
\begin{bmatrix}
x_1^{-2/3} & 0 \\ 0 & - x_2^{-2/3}
\end{bmatrix}
\delta x \le 0.
\end{align*}
Note that this needs to hold for all~$u \in \bR$. If we choose~$\delta x = [1,0]^\top$, then the following is required to hold for all~$x \in \bR^2$,
\begin{align*}
\frac{\partial M_{1,1}(x)}{\partial x_1}  - \frac{x_2^{1/3}}{x_1^{1/3}} \frac{\partial M_{1,1}(x)}{\partial x_2}  = - \frac{2}{3x_1} M_{1,1}(x) .
\end{align*}
We solve this partial differential equation by using the method of characteristics. Consider the following set of differential equations,
\begin{align*}
\frac{d x_2}{dx_1} = - \frac{x_2^{1/3}}{x_1^{1/3}}, \;
\frac{d M_{1,1}}{dx_1} =  - \frac{2}{3 x_1} M_{1,1}.
\end{align*}
The solution to the first and second equations are~$x_1^{2/3} + x_2^{2/3} = c_1$, i.e.,~$ x_2 = \pm (c_1-x_1^{2/3})^{3/2}$ and~$M_{1,1} = c_2 x_1^{-2/3}$, respectively, with integration constants~$c_1,c_2 \in \bR$. Therefore, we have
\begin{align*}
M_{1,1} \left(x_1, \pm \left(c_1-x_1^{2/3} \right)^{3/2} \right) = c_2 x_1^{-2/3}.
\end{align*}
This is not defined at~$x_1=0$ unless~$c_2 = 0$. If~$c_2=0$, then~$M(x)$ is not positive definite at~$x_2=\pm (c_1-x_1^{2/3})^{3/2}$. Therefore, the system is not differentially passive with respect to a positive definite storage function for any choice of the output function.  \red
\end{secex}

One notices that Krasovskii passivity can be viewed as the standard passivity for the extended system~\eqref{eq:esys} at a shifted equilibrium point~$(x^*,u^*) \in \cE$. Therefore, it is possible to develop control methodologies for the extended system based on Krasovskii passivity, which will be investigated in Section~\ref{sec:PBC} based on the following necessary and sufficient condition for Krasovskii passivity. 
\begin{secprop}\label{prop:KP}
Suppose that Assumption~\ref{asm:steady-state} holds. A system~\eqref{eq:sys} is Krasovskii passive at~$(x^*,u^*) \in \cE$ if and only if there exists a class~$C^1$ function~$S_K:\bR^n \times \bR^m \to \bR_+$ such that~$S_K(x^*,u^*)=0$ and
\begin{align}
&\frac{\partial S_K(x,u)}{\partial x} f(x,u)  \le 0, \label{cond1:KP}\\
&\frac{\partial S_K(x,u)}{\partial u} = h_K^\top (x,u)  \label{cond2:KP}
\end{align}
for all~$(x,u) \in \bR^{n\times m}$.
\red
\end{secprop}
\begin{IEEEproof}
The proof can be shown in a similar manner as that of Proposition~\ref{prop:passivity}.
\end{IEEEproof}

\subsection{Generalized Incremental Passivity and Shifted Passivity}\label{sec:IPSP}
In contraction analysis, differential properties have strong connections with the corresponding incremental properties such as stability \cite{FS:14,KBC:20,TRK:18,AJP:16}. Motivated by these analysis, we also consider incremental passivity, which is defined by using the following auxiliary system:
\begin{align}\label{eq:asys}
\left\{\begin{array}{l}
\dot x=f(x,u),\\
\dot x'=f(x',u'),\\
y_I = h_I(x,x'),
\end{array}\right.
\end{align}
where~$x':\bR_+ \to \bR^n$,~$u': \bR_+ \to \bR^m$,~$y_I:\bR_+ \to \bR^m$, and~$h_I : \bR^n \times \bR^n \to \bR^m$ is continuous and satisfies~$h_I(x,x)=0$ for all~$x \in \bR^n$. 

Incremental passivity is defined as a property of the auxiliary system by restricting~$h_I(x,x')$ into an incremental function~$h(x) - h(x')$ of some function~$h:\bR^n\to\bR^m$~\cite{PM:08,Simpson:18}. By releasing this restriction, we generalize the concept of incremental passivity 
\begin{secdefn}\label{def:IS}
The system~\eqref{eq:sys} is said to be \emph{generalized incrementally passive} if there exists a class~$C^1$ function~$S_I:\bR^n\times \bR^n\to \bR_+$ such that~$S_I(x,x)=0$ and
\begin{align}
&\frac{\partial S_I(x,x')}{\partial x}f(x,u) + \frac{\partial S_I(x,x')}{\partial x'}f(x',u') \nonumber\\
& \le (u - u')^\top h_I(x,x') \label{eq:IS}
\end{align}
for all~$(x,u) \in \bR^n \times \bR^m$~and $(x',u') \times \bR^n \times \bR^m$. \red
\end{secdefn}

As shown in the previous subsection, differential passivity implies Krasovskii passivity. As a counterpart, we have a similar relation between incremental and shifted passivity, where shifted passivity is introduced by substituting~$(x^*,u^*) \in \cE$ into~$(x',u')$ of Definition~\ref{def:IS} for incremental passivity.
\begin{secdefn}
Suppose that Assumption~\ref{asm:steady-state} holds. Then, the system~\eqref{eq:sys} is said to be \emph{shifted passive} at~$(x^*,u^*) \in \cE$ if there exists a class~$C^1$ function~$S_s:\bR^n \to \bR_+$ such that~$S_s(x^*)=0$ and
\begin{align*}
\frac{\partial S_s(x)}{\partial x}f(x,u)  \le (u - u^*)^\top (h(x) - h(x^*))
\end{align*}
for all~$(x,u) \in \bR^n \times \bR^m$. \red
\end{secdefn}

From their definitions, incremental passivity implies shifted passivity. 
\begin{secprop}
Under Assumption~\ref{asm:steady-state}, if a system~\eqref{eq:sys} is generalized incrementally passive, then it is shifted passive at any~$(x^*,u^*) \in \cE$ for~$h(x) = h_I(x,x^*)$. \red
\end{secprop}
\begin{IEEEproof}
This can be shown by substituting~$(x',u')=(x^*,u^*)$ into~\eqref{eq:IS}, where~$h_I(x^*,x^*)=0$ is used.
\end{IEEEproof}

Standard passivity is nothing but shifted passivity at~$(0,0)$ when~$h(0)=0$. That is, shifted passivity is defined by shifting an equilibrium point from~$(0,0)$ to an arbitrary~$(x^*,u^*) \in \cE$. Therefore its necessary and sufficient condition can readily be obtained by the slight modification of Proposition~\ref{prop:passivity}. Due to limitations of space, we refer to a similar result in \cite[Proposition 1]{JOG:07}. 

\subsection{Krasovskii Passivity and Shifted Passivity}
In Section~\ref{sec:DPKP}, we show that differential passivity implies Krasovskii passivity. Then, in Section~\ref{sec:IPSP}, we related in a similar way  generalized incremental passivity and shifted passivity. In this subsection, we add one additional relation, i.e., we show that  Krasovskii passivity implies shifted passivity.

\begin{secthm}\label{thm:KPSP}
Under Assumption~\ref{asm:steady-state}, if the system~\eqref{eq:sys} is Krasovskii passive at~$(x^*,u^*) \in \cE$, then it is shifted passive at~$(x^*,u^*)$ for
\begin{align}
h(x) = \left ( \frac{\partial S_K(x,u^*)}{\partial x} g (x) \right)^\top.
\label{KPSP_output}
\end{align}
\end{secthm}
\begin{IEEEproof}
By using Proposition~\ref{prop:KP}, we show that~$S_S(x)=S_K(x,u^*)$ is a storage function for shifted passivity. First,~$S_S(x^*)=S_K(x^*,u^*)=0$. Next, from~\eqref{eq:sys} and~\eqref{cond1:KP}, it follows that
\begin{align*}
&\frac{\partial S_S(x)}{\partial x} f(x,u)\\
&=\frac{\partial S_K(x,u^*)}{\partial x} f(x,u)\\
&=\frac{\partial S_K(x,u^*)}{\partial x} f(x,u^*) + \frac{\partial S_K(x,u^*)}{\partial x} g(x) (u-u^*)\\
&\le (u-u^*)^\top  \left ( \frac{\partial S_K(x,u^*)}{\partial x} g (x) \right)^\top.
\end{align*}
Note that~$S_K(x^*,u^*)=0$ and~$S_K (x,u) \ge 0$ for all~$(x,u) \in \bR^n \times \bR^m$ imply that~$S_K$ takes the minimum value at~$(x^*,u^*)$, and consequently
\begin{align*}
\frac{\partial S(x^*,u^*)}{\partial (x,u)}=0 \; \implies \; h(x^*)= \left ( \frac{\partial S_K(x^*,u^*)}{\partial x} g (x^*) \right)^\top = 0.
\end{align*}
Therefore, the system is shifted passive for $h(x)$ in~\eqref{KPSP_output}.
\end{IEEEproof}

As mentioned in Section~\ref{sec:DPKP}, we will develop a PBC technique on the basis of Krasovskii passivity in Section~\ref{sec:PBC}. Theorem~\ref{thm:KPSP} suggests that the developed technique can be modified for shifted passivity. We will also investigate this.

\subsection{Differential Passivity and Incremental Passivity}
To complete the relations between the various passivity concepts as in Fig.~\ref{fig:passivity},  in this subsection  we investigate a connection between differential passivity and incremental passivity.
\begin{secthm}
If a system~\eqref{eq:sys} is differentially passive with a storage function $S_D(x,\delta x) = (\delta x^\top M \delta x)/2$ for constant~$M \succeq 0$, then it is generalized incrementally passive for
\begin{align}
h_I(x,x') = \int_0^1 g^\top (s x + (1-s) x')M (x - x') ds.
\label{output:DSIS}
\end{align}
\end{secthm}
\begin{IEEEproof}
Let denote~$\gamma (s):= s x + (1-s) x'$ and~$\mu (s) := s u + (1-s) u',$ for $s \in [0,1]$. Compute
\begin{align*}
&f(x,u) - f(x',u') \\
&=\int_0^1 \frac{d f (\gamma (s),\mu(s))}{d s}  ds \\
&=\int_0^1 \left( F (\gamma (s),\mu(s)) \frac{d \gamma (s)}{d s} + g (\gamma (s)) \frac{d \mu (s)}{d s}\right) ds,
\end{align*}
where~$d \gamma (s)/d s = x - x'$ and~$d \mu (s)/d s = u - u'$.

By using this, we show that
\begin{align*}
S_I(x,x')=S_D\left(x, \frac{d\gamma(s)}{ds} \right) = \frac{1}{2} (x -x')^\top M (x -x')
\end{align*}
is a storage function for generalized incremental passivity. Since~$M$ is constant,~$\partial S_D(x,\delta x)/\partial x = 0$. From~\eqref{cond:DP} with~$\partial S_D(x,\delta x)/\partial x = 0$, it follows that
\begin{align*}
&\frac{\partial S_I(x,x')}{\partial x}f(x,u) + \frac{\partial S_I(x,x')}{\partial x'}f(x',u')\\
&=(x -x')^\top M (f(x,u) - f(x',u'))\\
&=\int_0^1 \frac{\partial S_D}{\partial \delta x} \left( F (\gamma (s),\mu(s)) \frac{d \gamma (s)}{d s} + g (\gamma (s)) \frac{d \mu (s)}{d s}\right) ds\\
& \le \int_0^1 \frac{\partial S_D}{\partial \delta x} g (\gamma (s)) \frac{d \mu (s)}{d s} ds\\
& \le (u - u')^\top  \int_0^1 g^\top (\gamma (s))M (x - x') ds, 
\end{align*}
where the arguments of~$S_D$ are~$(x,d\gamma (s)/ds)$. Therefore, the system is generalized incrementally passive for~$h_I(x,x')$ in~\eqref{output:DSIS}, where~$h_I(x,x)=0$ for all~$x \in \bR^n$.
\end{IEEEproof}

In the proof of the above theorem, we consider the straight line as a path connecting~$x$ and~$x'$ or~$u$ and~$u'$. One can however use an arbitrary class $C^1$ path. The integral in~\eqref{output:DSIS} depends on the considered path in general. 

As well known from~\cite{BCS:12} if $g_i^\top (x)Mdx$,~$i=1,\dots,m$ is an exact differential one-form, i.e., there exists a function~$h_i:\bR^n \to \bR$ such that 
\begin{align}
g_i^\top (x)M = \frac{\partial h_i(x)}{\partial x}, \label{eq:integrability}
\end{align}
then the path integral does not depend on the choice of a path. In the exact case,~$h_I$ becomes
\begin{align*}
h_I(x,x') = \begin{bmatrix} h_1(x)-h_1(x') & \cdots & h_m(x)-h_m(x') \end{bmatrix}^\top,
\end{align*}
and our incremental passivity matches the incremental passivity in literature~\cite{PM:08}. 

Moreover,~$h_i(x)$ is a linear function. The partial derivatives of both sides of~\eqref{eq:integrability} with respect to~$x$ yields
\begin{align*}
\frac{\partial^2 h_i(x)}{\partial x^2} = \frac{\partial^\top g_i(x)}{\partial x} M = \frac{1}{2} \left( \frac{\partial^\top g_i(x)}{\partial x} M + M \frac{\partial g_i(x)}{\partial x} \right),
\end{align*}
where we use the fact that~$\partial^2 h_i(x)/\partial x^2$ is symmetric. The inequality~\eqref{cond:DP} with~$S_D(x,\delta x) = (\delta x^\top M \delta x)/2$ implies~$\partial^2 h_i(x)/\partial x^2=0$. Therefore,~$h_i(x)$ can be described as~$c_i^\top x$ with~$c_i \in \bR^n$. Furthermore, one notices that~$g_i(x)$,~$i=1,\dots,m$ is also constant  if~$M \succ 0$.
Indeed, from~\eqref{eq:integrability}, it follows that~$g_i^\top =c_i^\top M^{-1}$.

\section{Passivity based Dynamic Controller Designs}\label{sec:PBC}
\subsection{Krasovskii Passivity based Controllers}
As mentioned, differential passivity does not provide a controller design method. However, we have proved in Section~\ref{sec:DPKP} that differential passivity implies Krasovskii passivity. In this section, we illustrate the utility of Krasovskii passivity for controller design.

For a Krasovskii passive system, we provide the following dynamic controller. 
\begin{secthm}\label{thm:KPcon}
Suppose that Assumption~\ref{asm:steady-state} holds, and the extended system~\eqref{eq:esys} is Krasovskii passive at $(x^*,u^*) \in \cE$ with respect to a storage function~$S_K(x,u)$. Consider the following dynamic controller for the extended system:
\begin{align}\label{eq:KPcon}
K_1\dot u_K= \nu_1 - K_2 u_K -K_3 (u-u^*) - y_K,
\end{align}
where~$\nu_1:\bR \to \bR^m$, and~$K_1 \succ 0$ and~$K_2,K_3 \succeq 0$ are free tuning parameters. Then, the following three statements hold:
\begin{itemize}
\item[(a)] The closed-loop system consisting of~\eqref{eq:esys} and~\eqref{eq:KPcon} is passive with respect to the supply rate~$\nu_1^\top u_K$.

\item[(b)] Let~$\nu_1=0$. Define
\begin{align}\label{eq:KPoutput}
\left\{\begin{array}{l}
\displaystyle y_1 := \frac{\partial S_K(x,u)}{\partial x} f(x,u),\\[2mm]
\displaystyle y_2 := K_3 (u-u^*) + y_K.
\end{array}\right.
\end{align}
Suppose that there exists~$r>0$ such that
\begin{align*}
S_K(x,u) + |u-u^*|_{K_3}^2/2 > 0
\end{align*}
for all~$(x,u) \in B_r(x^*,u^*)\setminus \{(x^*,u^*)\}$, where~$B_r$ is an open ball as defined in the notation part. Then, there exists~$\bar r>0$ such that any solution to the closed-loop system starting from~$B_{\bar r}(x^*,u^*,0)$ converges to the largest invariant set contained in
\begin{align}\label{eq:invset_KPcon}
\{ (x,u,u_K) &\in B_{\bar r}(x^*,u^*,0) : y_1=0, K_2 u_K = 0 \}.
\end{align}
\item[(c)] Moreover, if~$K_2 \succ 0$ and the extended system with the outputs~$(y_1,y_2)$ is detectable at~$((x^*,u^*),0)$, then~$(x^*,u^*)$ is asymptotically stable. 
\end{itemize}
\end{secthm}
\begin{IEEEproof}
Consider the following storage function:
\begin{align}
S_1(x,u,u_K):=S_K(x,u) + \frac{1}{2} |u-u^*|_{K_3}^2 + \frac{1}{2} |u_K|_{K_1}^2,
\label{eq:S1}
\end{align}
which is positive semi-definite at~$(x^*,u^*,0)$. By using~\eqref{eq:esys},~\eqref{cond2:KP} and~\eqref{eq:KPcon}, the Lie derivative of~$S_1$ along the vector field of the closed-loop system, simply denoted by~$dS_1/dt$, is computed as
\begin{align*}
\frac{dS_1}{dt}
=& \frac{\partial S_K(x,u)}{\partial x} f(x,u)\\
& + u_K^\top \left( \frac{\partial^\top S_K(x,u)}{\partial u} + K_1 \dot u_K +K_3 (u-u^*) \right) \\
=& \frac{\partial S_K(x,u)}{\partial x} f(x,u) + u_K^\top(\nu_1 - K_2 u_K).
\end{align*}
Therefore, (a) follows from~\eqref{cond1:KP}.

Next, we show~(b). From the assumption,~$S_1(x,u,u_K)$ is positive definite at~$(x^*,u^*,0)$ in~$B_r(x^*,u^*) \times \bR^m$. Therefore, the statement follows from LaSalle's invariance principle, where recall~\eqref{eq:KPoutput}. Finally, one can show~(c) in a similar manner as~\cite[Corollary 4.2.2]{Schaft:00}, where~$y_2$ comes from~\eqref{eq:KPcon} with~$u_K(\cdot) =0$.
\end{IEEEproof}

The controller can be interpreted in terms of the transfer function. If~$K_1s^2 + K_2 s + K_3$ is invertible, the Laplace transformation of the controller dynamics~\eqref{eq:KPcon} can be computed as
\begin{align}\label{eq:KPcon_Laplace}
U(s) = (K_1s^2 + K_2 s + K_3 )^{-1} (V_1(s) - Y_K(s)),
\end{align}
where~$U(s)$,~$Y_K(s)$,~$V_1(s)$ denote the Laplace transformations of~$u-u^*$,~$y_K$ and~$\nu_1$, respectively. Therefore, the controller can be viewed as a second order output feedback controller.

In the above theorem,~$K_1$ is chosen to be positive definite. However, even if~$K_1=0$, by assuming positive definiteness of~$K_2$ instead, it is possible to construct a first order controller. The following corollary is a generalization of a controller for boost converters in DC microgrids~\cite{KCS:18,CLK:19} to general nonlinear systems. The proof is similar to the proof that can be found in~\cite{KKS:19} and is omitted.
\begin{seccor}
Suppose that the assumptions in Theorem~\ref{thm:KPcon} hold. Consider the extended system~\eqref{eq:esys} with the following controller:
\begin{align}\label{eq:Krasovskii_passivity_controller2}
K_2 u_K = \nu_1 -K_3 (u-u^*) - y_K,
\end{align}
where~$\nu_1:\bR \to \mathbb{R}^m$, and~$K_2 \succ 0$ and~$K_3 \succeq 0$ are free tuning parameters. Then, the following three statements hold:
\begin{itemize}
\item[(a)] The closed-loop system consisting of~\eqref{eq:esys} and~\eqref{eq:Krasovskii_passivity_controller2} is passive with respect to the supply rate~$\nu_1^\top u_K$.

\item[(b)]Let $\nu_1=0$. Suppose that there exists~$r>0$ such that
\begin{align*}
S_K(x,u) + |u-u^*|_{K_3}^2/2 > 0
\end{align*}
for all~$(x,u) \in B_r(x^*,u^*)\setminus \{(x^*,u^*)\}$. Then, there exists~$\bar r>0$ such that any solution to the closed-loop system starting from~$B_{\bar r}(x^*,u^*)$ converges to the largest invariant set contained in
\begin{align*}
\{ (x,u) \in B_{\bar r}(x^*,u^*) : y_1=0, y_2 =0 \}
\end{align*}
for~$(y_1,y_2)$ in~\eqref{eq:KPoutput}.
\item[(c)] Moreover, if the extended system with the outputs~$(y_1, y_2)$ is detectable at~$((x^*,u^*),0)$, then~$(x^*,u^*)$ is asymptotically stable.
\red
\end{itemize}
\end{seccor}

\subsection{Shifted Passivity based Controllers}
Theorem~\ref{thm:KPSP} shows that Krasovskii passivity implies shifted passivity. Inspired by this fact and Krasovskii PBCs proposed in the previous subsection, we provide a shifted passivity based controller for the original system~\eqref{eq:sys} instead of its extended system.
\begin{secthm}\label{thm:SPcon}
Suppose that the assumptions in Theorem~\ref{thm:KPcon} hold. For the original system~\eqref{eq:sys} with the output function in~\eqref{KPSP_output}, consider the following dynamic feedback controller:	
\begin{align}\label{eq:SPcon}
\left\{\begin{array}{l}
\displaystyle \hspace{4mm}u = u^* - K_5 y + K_6 v,\\[2mm]
\displaystyle K_4 \dot v = \nu_2 - K_6 y - K_7 v,
\end{array}\right.
\end{align}
where $\nu_2:\bR \to \bR^m$, and~$K_4 \succ 0$ and~$K_5,K_6,K_7 \succeq 0$ are free tuning parameters. Then, the following three statements hold:
\begin{itemize}
\item[(a)] The closed-loop system consisting of~\eqref{eq:sys}  and~\eqref{eq:SPcon} is passive with respect to the supply rate~$\nu_2^\top v$.

\item[(b)] Let~$\nu_2=0$. Define
\begin{align}\label{eq:KPoutput2}
y_3 := \frac{\partial S_K(x,u^*)}{\partial x} f(x,u^*).
\end{align}
Suppose that there exists~$r>0$ such that~$S_K(x,u^*)>0$ for all~$x \in B_r(x^*)\setminus \{x^*\}$. Then, there exists~$\bar r>0$ such that any solution to the closed-loop system starting from~$B_{\bar r}(x^*,0)$ converges to the largest invariant set contained in
\begin{align}\label{eq:invset_SPcon}
\{ (x,v) \in B_{\bar r}(x^*,0) : K_5 y =0, y_3 = 0, K_7 v = 0 \}.
\end{align}
\item[(c)] Moreover, if~$K_5,K_7 \succ 0$ and the system with the outputs~$(y,y_3)$ is detectable at~$(x^*,u^*)$, then~$(x^*,u^*)$ is asymptotically stable.
\end{itemize}
\end{secthm}
\begin{IEEEproof}
Consider the following storage function:
\begin{align}
S_2(x,v):= S_K(x,u^*) +\dfrac{1}{2} | v|_{K_4}^2,
\label{eq:S2}
\end{align}
which is positive semi-definite at~$(x^*,0)$.
In a similar manner as the proof of Theorem~\ref{thm:KPSP}, by using~\eqref{KPSP_output} and~\eqref{eq:SPcon}, the Lie derivative of~$S_2$ along the vector field of the closed-loop system, simply denoted by $dS_2/dt$, is computed as follows
\begin{align*}
\frac{dS_2}{dt}=&\; \frac{\partial S_K(x,u^*)}{\partial x} f(x,u^*) + \frac{\partial S_K(x,u^*)}{\partial x}g(x) (u-u^*)\\
&+v^\top K_4 \dot v\\
=&\; v^\top \nu_2 + \frac{\partial S_K(x,u^*)}{\partial x} f(x,u^*)  - v^\top K_7 v - y^\top K_5 y.
\end{align*}
From~\eqref{cond1:KP}, we obtain~(a). Also, one can show~(b) and (c) from~\eqref{eq:KPoutput2} in similar manners as the proof of Theorem~\ref{thm:KPcon}.
\end{IEEEproof}
\begin{secrem}\label{rem:SPcon}
Similar conclusions as Theorem~\ref{thm:SPcon} hold if Assumption~\ref{asm:steady-state} holds, and the original system~\eqref{eq:sys} is shifted passive with respect to a storage function~$S_S(x)$. 
\red
\end{secrem}

We again interpret the proposed controller in terms of the transfer function. If~$K_4 s  + K_7$ is invertible, the Laplace transformation of the controller dynamics~\eqref{eq:SPcon} can be computed as
\begin{align*}
U(s) =  - K_5 Y(s) + K_6 (K_4 s  + K_7)^{-1} (V_2(s) - K_6 Y(s)).
\end{align*}
where~$U(s)$,~$Y(s)$,~$V_2(s)$ denote the Laplace transformations of~$u-u^*$,~$y$ and~$\nu_2$, respectively. Therefore, the controller can be viewed as a proper output feedback controller and is different from~\eqref{eq:KPcon_Laplace}. If $K_5=0$, one has a structure of the low pass filter. If $K_6=0$, one has a standard-type PBC. If~$K_7=0$, one has a PI feedback controller, which is an extension of one presented in~\cite{JOG:07}. 

\section{Example}\label{sec:Ex}
\begin{figure}[tb]
\begin{center}
\includegraphics[width=80mm]{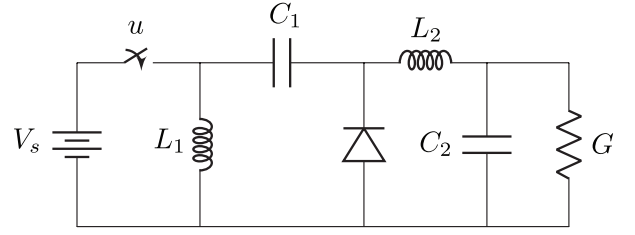}
\caption{Electrical scheme of the Zeta converter.}
\label{fig:zeta}
\end{center}
\end{figure}

In this example, we consider the average model of a DC-Zeta converter. It has the capability of both buck and boost converters, i.e., it can amplify and reduce the supply voltage while maintaining the polarity. The schematic of the Zeta converter is given in Fig.~\ref{fig:zeta}. As shown, it contains four energy storage elements, namely two inductors~$L_1$,~$L_2$ and two capacitors~$C_1$,~$C_2$, an ideal switching element~$u$ and an ideal diode. Further,~$V_s$ and~$G$ denote the constant supply voltage and the load, respectively. The objective of the converter is to maintain a desired voltage~$v^*$ across the load~$G$. After some changes of state and time variables, one obtains the following normalized model for the converter; for more details about changes of variables, see~\cite[Chapter~2.8]{SS:06}.

\begin{align}\label{eq:zeta}
\dot x = 	
\begin{bmatrix}
-x_2 \\ x_1 \\ - x_4/\alpha_1 \\ (x_3 - x_4/\alpha_3 )/\alpha_2
\end{bmatrix}
+\begin{bmatrix}
1+x_2 \\ -(x_1+x_3) \\ (1+x_2)/\alpha_1 \\ 0
\end{bmatrix}u,
\end{align}
where $\alpha_1$, $\alpha_2$ and $\alpha_3$ are positive constants depending on the system parameters. {It is worth pointing out that a (standard) PBC has not been provided for this class of systems because it is difficult to find a storage function. However, we demonstrate that our proposed two PBC techniques are useful for controller design.}

For this system, the set~$\cE$ is obtained as
\begin{align*}
\cE_v=\biggl\{&(x^*,u^*)  \in \bR^4 \times \bR:\\
& x^*=\left(\dfrac{(v^\ast)^2}{\alpha_3},v^\ast,\dfrac{v^\ast}{\alpha_3},v^\ast\right),  u^*= \dfrac{v^\ast}{v^\ast+1},\biggr\}, \; v^\ast \in \bR_+.
\end{align*} 
One notices that~$\cE_v$ has a unique element for any~$v^\ast \in \bR_+$. The parameters are chosen as~$a_1=a_2=a_3=1$ and~$v^*=1/3$ which determines~$x^*=[1/9, 1/3, 1/3, 1/3]$ and~$u^*=1/4$. 

First, we illustrate the Krasovskii PBC~\eqref{eq:KPcon} in Theorem~\ref{thm:KPcon}. One can confirm that the DC-Zeta converter~\eqref{eq:zeta} satisfies~\eqref{cond:DP} for
\begin{align}
S_K(x,u) = |f(x,u)|_M/2, \; M={\rm diag}\{1,1,\alpha_1,\alpha_2\},
\label{eq:zeta_storage}
\end{align}
and the following holds.
\begin{align*}
h_K(x,u) &=\frac{\partial S_K(x,u)}{\partial u}\\
& =  
\begin{bmatrix}
1+x_2 & -(x_1+x_3) & 1+x_2 & 0
\end{bmatrix} \dot x.
\end{align*}
If~$x$ and~$\dot x$ are measured, the controller~\eqref{eq:KPcon} does not require the information of parameters~$\alpha_1,\alpha_2,\alpha_3$ for the DC-Zeta converter. 

\begin{figure}[tb]
\centering
\includegraphics[width=\linewidth]{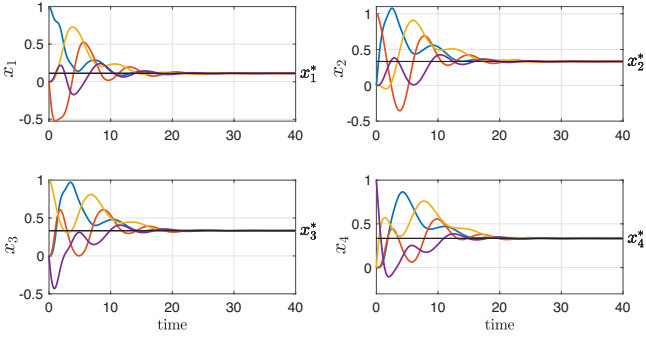}
\caption{Closed-loop trajectories when controlled by the Krasovskii PBC.}
\label{fig:KPsim}
\end{figure}

It is possible to show that the storage function~$S_1(x,u,u_K)$ in~\eqref{eq:S1} is radially unbounded, and the largest invariant set contained in the set~\eqref{eq:invset_KPcon} is nothing but~$\cE_v$. Therefore, for any~$v^* \in \bR_+$, any trajectory of the closed-loop system converges to~$\cE_v$. Figure~\ref{fig:KPsim} shows the trajectories of the closed-loop system starting from several initial states, where~$\nu_1=0$ and~$K_1=K_2=K_3=1$.

\begin{figure}[tb]
\includegraphics[width=\linewidth]{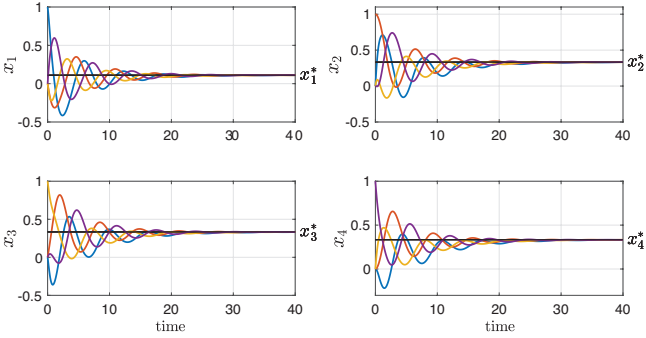}
\caption{Closed-loop trajectories when controlled by the shifted PBC.}
\label{fig:SPsim}
\end{figure}

Second, we illustrate the shifted PBC~\eqref{eq:SPcon} in Theorem~\ref{thm:SPcon}, where
\begin{align*}
h(x)&= \left( \frac{\partial S_K(x,u^*)}{\partial x} g(x) \right)^\top \\
&= 
\frac{1}{2}
\begin{bmatrix}
(1- u^*) (x_1 + x_2) \\
(1 - u^* +u^*/\alpha_1) (1+x_2)\\
- u^* (x_1 + x_2)\\
(1/\alpha_1)(1+ x_2)
\end{bmatrix}^\top f(x,u^*).
\end{align*}
Again, it is possible to show that  the storage function~$S_2(x,v)$ in~\eqref{eq:S2} is radially unbounded, and the largest invariant set contained in the set~\eqref{eq:invset_SPcon} is nothing but~$\cE_v$. Therefore, for any~$v^* \in \bR_+$, any trajectory of the closed-loop system converges to~$\cE_v$. Figure~\ref{fig:SPsim} shows the trajectories of the closed-loop system starting from several initial states, where~$\nu_2=0$ and~$K_4=K_5=K_6=K_7=1$. 


Finally, simulation results indicate that the Krasovskii PBC achieves convergence to~$\cE_v$ with less  oscillations than the shifted PBC, whereas the shifted PBC converges with relatively lower amplitudes of the oscillations than the Krasovskii PBC case.
 
\section{Conclusion}\label{sec:con}
In this paper, we have introduced the concept of Krasovskii passivity for addressing the difficulty of differential passivity based controller design. First, we have established relations among the relevant four passivity concepts: 1) differential passivity, 2) Krasovskii passivity, 3) generalized incremental passivity, and 4) shifted passivity. Next, we have proposed Krasovskii/shifted passivity based dynamic controllers. The utility of the proposed controllers has been illustrated by the DC-Zeta converter. Future work includes studying control methodologies for networked Krasovskii passive systems as done for shifted passivity-short systems~\cite{SZ:19}.

\bibliographystyle{IEEEtran}
\bibliography{reference}

\begin{thebibliography}{10}
\providecommand{\url}[1]{#1}
\csname url@samestyle\endcsname
\providecommand{\newblock}{\relax}
\providecommand{\bibinfo}[2]{#2}
\providecommand{\BIBentrySTDinterwordspacing}{\spaceskip=0pt\relax}
\providecommand{\BIBentryALTinterwordstretchfactor}{4}
\providecommand{\BIBentryALTinterwordspacing}{\spaceskip=\fontdimen2\font plus
\BIBentryALTinterwordstretchfactor\fontdimen3\font minus
  \fontdimen4\font\relax}
\providecommand{\BIBforeignlanguage}[2]{{%
\expandafter\ifx\csname l@#1\endcsname\relax
\typeout{** WARNING: IEEEtran.bst: No hyphenation pattern has been}%
\typeout{** loaded for the language `#1'. Using the pattern for}%
\typeout{** the default language instead.}%
\else
\language=\csname l@#1\endcsname
\fi
#2}}
\providecommand{\BIBdecl}{\relax}
\BIBdecl

\bibitem{KKS:19}
K.~C. Kosaraju, Y.~Kawano, and J.~M.~A. Scherpen, ``Krasovskii's passivity,''
  \emph{IFAC-PapersOnLine}, vol.~52, no.~16, pp. 466--471, 2019.

\bibitem{Angeli:02}
D.~Angeli, ``A {L}yapunov approach to incremental stability properties,''
  \emph{IEEE Transactions on Automatic Control}, vol.~47, no.~3, pp. 410--421,
  2002.

\bibitem{LS:98}
W.~Lohmiller and J.~J.~E. Slotine, ``On contraction analysis for non-linear
  systems,'' \emph{Automatica}, vol.~34, no.~6, pp. 683--696, 1998.

\bibitem{FS:14}
F.~Forni and R.~Sepulchre, ``A differential {L}yapunov framework for
  contraction analysis,'' \emph{IEEE Transactions on Automatic Control},
  vol.~59, no.~3, pp. 614--628, 2014.

\bibitem{KBC:20}
Y.~Kawano, B.~Besselink, and M.~Cao, ``Contraction analysis of monotone systems
  via separable functions,'' \emph{IEEE Transactions on Automatic Control},
  vol.~65, no.~8, pp. 3486--3501, 2020.

\bibitem{TRK:18}
D.~N. Tran, B.~S. R{\"u}ffer, and C.~M. Kellett, ``Convergence properties for
  discrete-time nonlinear systems,'' \emph{IEEE Transactions on Automatic
  Control}, vol.~64, no.~8, pp. 3415--3422, 2018.

\bibitem{AJP:16}
V.~Andrieu, B.~Jayawardhana, and L.~Praly, ``Transverse exponential stability
  and applications,'' \emph{IEEE Transactions on Automatic Control}, vol.~61,
  no.~11, pp. 3396--3411, 2016.

\bibitem{KS:16}
Y.~Kawano and J.~M.~A. Scherpen, ``Model reduction by differential balancing
  based on nonlinear {H}ankel operators,'' \emph{IEEE Transactions on Automatic
  Control}, vol.~62, no.~7, pp. 3293--3308, 2016.

\bibitem{PM:08}
A.~Pavlov and L.~Marconi, ``Incremental passivity and output regulation,''
  \emph{Systems \& Control Letters}, vol.~57, no.~5, pp. 400 -- 409, 2008.

\bibitem{Schaft:13}
A.~J. van~der Schaft, ``On differential passivity,'' \emph{IFAC Proceedings
  Volumes}, vol.~46, no.~23, pp. 21 -- 25, 2013.

\bibitem{FSS:13}
F.~Forni, R.~Sepulchre, and A.~J. van~der Schaft, ``On differential passivity
  of physical systems,'' \emph{Proc. the 52nd IEEE Conference on Decision and
  Control}, pp. 6580--6585, 2013.

\bibitem{Schaft:00}
A.~J. van~der Schaft, \emph{{$L_2$}-Gain and Passivity Techniques in Nonlinear
  Control}.\hskip 1em plus 0.5em minus 0.4em\relax Springer, 2000, vol.~2.

\bibitem{KCS:18}
K.~C. Kosaraju, M.~Cucuzzella, J.~M.~A. Scherpen, and R.~Pasumarthy,
  ``Differentiation and passivity for control of {B}rayton-{M}oser systems,''
  \emph{IEEE Transactions on Automatic Control}, 2020, (early access).

\bibitem{KCP:18}
K.~C. Kosaraju, V.~Chinde, R.~Pasumarthy, A.~Kelkar, and N.~M. Singh,
  ``Differential passivity like properties for a class of nonlinear systems,''
  \emph{Proc. 2018 Annual American Control Conference}, pp. 3621--3625, 2018.

\bibitem{CLK:19}
M.~Cucuzzella, R.~Lazzari, Y.~Kawano, K.~C. Kosaraju, and J.~M.~A. Scherpen,
  ``Robust passivity-based control of boost converters in {DC} microgrids,''
  \emph{Proc. 58th IEEE Conference on Decision and Control}, 2019.

\bibitem{RDS:18}
R.~Reyes-B{\'a}ez, A.~Donaire, A.~J. van~der Schaft, B.~Jayawardhana, and
  T.~Perez, ``Tracking control of marine craft in the port-{H}amiltonian
  framework: A virtual differential passivity approach,'' \emph{arXiv preprint
  arXiv:1803.07938}, 2018.

\bibitem{JOG:07}
B.~Jayawardhana, R.~Ortega, E.~Garc{\'i}a-Canseco, and F.~Casta{\~ n}os,
  ``Passivity of nonlinear incremental systems: Application to {PI}
  stabilization of nonlinear {RLC} circuits,'' \emph{Systems \& Control
  Letters}, vol.~56, no.~9, pp. 618 -- 622, 2007.

\bibitem{OSM:01}
R.~Ortega, A.~J. van~der Schaft, I.~Mareels, and B.~Maschke, ``Putting energy
  back in control,'' \emph{IEEE Control Systems Magazine}, vol.~21, no.~2, pp.
  18--33, 2001.

\bibitem{Simpson:18}
J.~W. Simpson-Porco, ``Equilibrium-independent dissipativity with quadratic
  supply rates,'' \emph{IEEE Transactions on Automatic Control}, vol.~64,
  no.~4, pp. 1440--1455, 2019.

\bibitem{Khalil:96}
H.~K. Khalil, \emph{Nonlinear Systems}.\hskip 1em plus 0.5em minus 0.4em\relax
  New Jersey: Prentice-Hall, 1996.

\bibitem{Terrell:09}
W.~J. Terrell, \emph{Stability and Stabilization: An Introduction}.\hskip 1em
  plus 0.5em minus 0.4em\relax Princeton University Press, 2009.

\bibitem{Schaft:82}
A.~J. van~der Schaft, ``Observability and controllability for smooth nonlinear
  systems,'' \emph{SIAM Journal on Control and Optimization}, vol.~20, no.~3,
  pp. 338--354, 1982.

\bibitem{BCS:12}
D.~Bao, S.~S. Chern, and Z.~Shen, \emph{An Introduction to Riemann-Finsler
  Geometry}.\hskip 1em plus 0.5em minus 0.4em\relax New York: Springer-Verlag,
  2012.

\bibitem{SS:06}
H.~J. Sira-Ramirez and R.~Silva-Ortigoza, \emph{Control Design Techniques in
  Power Electronics Devices}.\hskip 1em plus 0.5em minus 0.4em\relax Springer
  Science \& Business Media, 2006.

\bibitem{SZ:19}
M.~Sharf and D.~Zelazo, ``Network feedback passivation of passivity-short
  multi-agent systems,'' \emph{IEEE Control Systems Letters}, vol.~3, no.~3,
  pp. 607--612, 2019.

\end{thebibliography}
\end{document}